\documentstyle[sprocl]{article}

\def\Journal#1#2#3#4{{#1} {\bf #2}, #3 (#4)}

\def\NPB{{\em Nucl. Phys.} B}
\def\PLB{{\em Phys. Lett.}  B}

\def\PRD{{\em Phys. Rev.} D}
\def\ZPC{{\em Z. Phys.} C}

\def\TMP{\em Theor. Math. Phys.} 

\def\be{\begin{equation}}
\def\ee{\end{equation}}
\def\bea{\begin{eqnarray}}
\def\eea{\end{eqnarray}}

\newcommand{\half}{\mbox{\small{$\frac{1}{2}$}}} 
  
\newcommand{\Nf}{N_{\!f}} 
\newcommand{\Nff}{\tilde{N}_{\!f}} 
\newcommand{\MSbar}{\overline{\mbox{MS}}}

\begin{document}

\title{STATUS OF TWIST-$2$ OPERATOR DIMENSIONS AT $O(1/\Nf)$\footnote{Talk 
presented at Deep Inelastic Scattering 98, held at IIHE-ULB, Brussels, 
Belgium, 4th-8th April, 1998.} 
} 
\author{J.A. GRACEY} 
\address{Theoretical Physics Division, Department of Mathematical Sciences, 
\\ University of Liverpool, Peach Street, Liverpool, L69 7ZF, UK 
\\ E-mail: jag@amtp.liv.ac.uk}

\maketitle\abstracts{We review the computation of the anomalous dimensions of 
the twist-$2$ unpolarized operators in the large $\Nf$ expansion. Results are 
discussed for the predominantly gluonic singlet operator and the $O(1/\Nf)$ 
part of the $3$-loop splitting function is given.} 

{\vspace{-6cm} 
\hspace{9.8cm} 
LTH 425} 
{\vspace{5.5cm}} 

\section{Introduction} 
A current problem in multiloop perturbation theory is the construction of the 
$3$-loop terms of the twist-$2$ operator dimensions which appear in the 
operator product expansion used in deep inelastic scattering. The calculation 
of the $\MSbar$ coefficients as a function of the momentum fraction $x$ or 
equally the operator moment $n$, is necessary to perform the full $2$-loop 
coefficient function evolution using the renormalization group. Currently the 
full $2$-loop results as a function of $n$ are known for the twist-$2$ flavour 
non-singlet and singlet, unpolarized and polarized 
operators.$\,^{\mbox{\footnotesize{1-4}}}$ At three loops exact results are 
known for the first four even moments for the unpolarized case and in addition 
for the non-singlet $n$ $=$ $10$ moment.\cite{5} However, the full result as a 
function of $n$ has yet to be determined. As a first step in this direction, 
Matiounine et al have recently computed the finite parts of all the $2$-loop 
diagrams for the twist-$2$ operators.\cite{7} These are required as they will 
give contributions at $3$-loops when multiplied by $1$-loop counterterms. Aside
from the perturbative expansion one can gain insight into the structure of the 
dimensions in other approximations. For instance, a low-$x$ analysis can also 
be performed.\cite{8} 

Another expansion technique which has been applied to this problem is the 
$1/\Nf$ method where $\Nf$ is the number of quarks. In this reordering of 
perturbation theory, where chains of quark bubbles form the dominant 
contribution, one can probe the perturbative structure beyond currently known
orders. In particular results as a function of $n$ can be provided for the 
$3$-loop coefficients at $O(1/\Nf)$ as $\Nf$ $\rightarrow$ $\infty$ as well as 
at all higher loops.\cite{9,10} These have been important in verifying the 
correctness of the results in \cite{5,6} in the region of overlap. Currently 
the $1/\Nf$ method has been applied to the twist-$2$ non-singlet and singlet 
fermionic operators. More recently the anomalous dimension of the outstanding 
singlet gluonic operator has been given in \cite{11}, which we focus on here. 

\section{Formalism} 
In standard notation the basic twist-$2$ unpolarized singlet operators are, 
\begin{eqnarray}
{\cal O}^{\mu_1 \ldots \mu_n}_{\mbox{\footnotesize{$q$}}} &=&  
i^{n-1} {\cal S} \bar{\psi}^I \gamma^{\mu_1} D^{\mu_2} \ldots D^{\mu_n} 
\psi^I - \mbox{trace terms} \nonumber \\ 
{\cal O}^{\mu_1 \ldots \mu_n}_{\mbox{\footnotesize{$g$}}} &=&  
\half i^{n-2} {\cal S} \, \mbox{tr} \, G^{a \, \mu_1\nu} D^{\mu_2} 
\ldots D^{\mu_{n-1}} G^{a \,\, ~ \mu_n}_{~~\nu} - \mbox{trace terms}
\label{op1}
\end{eqnarray}
As the operators ${\cal O}_{\mbox{\footnotesize{$q$}}}$ and 
${\cal O}_{\mbox{\footnotesize{$g$}}}$ have the same canonical dimension they
will mix under renormalization.\cite{1} Hence one needs to introduce a 
mixing matrix, $\gamma_{ij}(a)$, of anomalous dimensions. The $\Nf$ dependence 
in the perturbative expansion of each entry in $\gamma_{ij}(a)$ is not the 
same. For example, with $\Nff$ $=$ $T(R)\Nf$ 
\begin{eqnarray}  
\gamma_{qq}(a) &=& a_1a + (a_{21}\Nff + a_{22})a^2 + (a_{31}\Nff^2 + a_{32}\Nff 
+ a_{33})a^3 + O(a^4) \nonumber \\ 
\gamma_{gq}(a) &=& b_1a + (b_{21}\Nff + b_{22})a^2 + (b_{31}\Nff^2 + b_{32}\Nff 
+ b_{33})a^3 + O(a^4) \nonumber \\ 
\gamma_{qg}(a) &=& c_1 \Nff a + c_2\Nff a^2 + (c_{31}\Nff^2 + c_{32}\Nff 
+ c_{33})a^3 + O(a^4) \nonumber \\ 
\gamma_{gg}(a) &=& (d_{11}\Nff + d_{12})a + (d_{21}\Nff + d_{22})a^2 
\nonumber \\ 
&& +~ (d_{31}\Nff^2 + d_{32}\Nff + d_{33})a^3 + O(a^4)  
\label{matdef} 
\end{eqnarray} 
In the $1/\Nf$ approach,\cite{9,10,11}, one computes sets of these 
coefficients by considering QCD at its non-trivial $d$-dimensional fixed point 
and studies the scaling behaviour of the appropriate Green's function there. 
For the present problem the resulting exponents give the eigen-anomalous 
dimensions of $\gamma_{ij}(a)$ at criticality. In terms of the perturbative 
coefficients the $O(1/\Nf)$ eigenoperator dimensions involve the combinations 
$(a_{l1}$ $-$ $b_{l1}c_1/d_{11})$ for ${\cal O}_{\mbox{\footnotesize{$q$}}}$ 
and $(d_{l1}$ $+$ $b_{l1}c_1/d_{11})$ for 
${\cal O}_{\mbox{\footnotesize{$g$}}}$ at the $l$-th loop.  

\section{Results} 
The application to QCD of the basic formalism developed in \cite{12} for simple
scalar theories yields an expression for the dimension of 
${\cal O}_{\mbox{\footnotesize{$g$}}}$ at $O(1/\Nf)$ as a function of $n$ and 
the space-time dimension $d$.\cite{11} The full all orders expression is given 
explicitly in \cite{11}. Its $\epsilon$-expansion, where $d$ $=$ $4$ $-$ 
$2\epsilon$, agrees with all previous perturbative 
calculations.\cite{1,2,3,6} At $3$-loops the numerical values of the leading 
order gluonic eigen-operator coefficients are given in Table 1 and these can be 
compared with the exact coefficients given in \cite{11}. Clearly the modulus of
these coefficients increases slowly with the moment. Another feature of the 
results is that since $b_{31}$ depends only on the colour Casimir $C_2(R)$, 
then the $\epsilon^3$ coefficient of $C_2(G)$ in the $\epsilon$-expansion of 
gluonic eigen-dimension gives the exact $3$-loop dependence of $d_{31}$ as a 
function of $n$. Hence we can determine the $x$-dependence of the gluonic DGLAP
splitting function which is proportional to $C_2(G)$. Using the Mellin 
transform we deduce  
\begin{eqnarray} 
P^{\footnotesize{\mbox{$3$-loop}}}_{gg}(x,C_2(G)) &=& -~ \frac{1}{54} \left[
87 \delta(1-x) ~+~ ( 304 + 172x + 208x^2)\ln x \right. \nonumber \\  
&&~~~~~~~~ \left. -~ 48(1+x) \ln^2 x ~+~ 32 ~-~ \frac{32}{[1-x]_+} \right. 
\nonumber \\ 
&&~~~~~~~~ \left. +~ 192(1+x) (\psi^\prime(1) - \mbox{Li}_2(x) ) \right. 
\nonumber \\ 
&&~~~~~~~~ \left. +~ \frac{4(1-x)}{x} (52 + 19x + 52x^2) \ln(1-x) \right. 
\nonumber \\
&&~~~~~~~~ \left. +~ \frac{4(1-x)}{3x}( 236 + 47x + 236x^2) \right] 
\end{eqnarray}  
where $\mbox{Li}_2(x)$ is the dilogarithm function and $\psi(x)$ is the 
derivative of the logarithm of the Euler $\Gamma$-function. 

{\begin{table}[t] 
\caption{Numerical values of the coefficients of $\left[ d_{31} \right.$ $+$ 
$\left. b_{31}c_1/d_{11} \right]$.} 
\begin{center} 
\begin{tabular}{|r||r|r|} 
\hline 
$n$ & $~~~C_2(R)$ coefficient & $~~~C_2(G)$ coefficient \\ 
\hline  
2 & $ - \, 11.1769547325$ & $ - \, 17.415637860$ \\ 
4 & $ - \, 6.1986353909$ & $ - \, 12.475078189$ \\   
6 & $ - \, 5.1270609536$ & $ - \, 12.665273968$ \\ 
8 & $ - \, 4.8386758731$ & $ - \, 13.094409108$ \\ 
10 & $ - \, 4.7463824737$ & $ - \, 13.507443429$ \\ 
12 & $ - \, 4.7180193885$ & $ - \, 13.876218903$ \\ 
14 & $ - \, 4.7136211552$ & $ - \, 14.202839253$ \\ 
16 & $ - \, 4.7187348148$ & $ - \, 14.493720966$ \\ 
18 & $ - \, 4.7275177819$ & $ - \, 14.754952547$ \\ 
20 & $ - \, 4.7374464091$ & $ - \, 14.991545068$ \\ 
22 & $ - \, 4.7473994042$ & $ - \, 15.207488437$ \\ 
24 & $ - \, 4.7568866098$ & $ - \, 15.405947998$ \\ 
\hline 
\end{tabular} 
\end{center} 
\end{table} } 

\section{Discussion} 
The provision of the gluonic operator dimension in \cite{11} now completes the 
$O(1/\Nf)$ examination of the twist-$2$ unpolarized operator dimensions. More 
recently the same calculation has been completed for the polarized 
operators.\cite{13} Future calculations in this area would involve computing 
the anomalous dimensions to the next order, $O(1/\Nf^2)$. The starting point 
for this would be the non-singlet sector. 

\section*{Acknowledgements} This work was carried out through a {\sc PPARC} 
Advanced Fellowship.

\end{document}